\documentclass[fleqn,twocolumn]{article}

\usepackage[cp1251]{inputenc}
\usepackage[english]{babel}

\usepackage[dvips]{graphicx}
\usepackage{amssymb}
\usepackage{amsmath}

\begin{document}

\renewcommand{\bibname}{\normalsize }
\renewcommand{\abstractname}{\normalsize \textbf Abstract}
\renewcommand{\thefootnote}{\alph{footnote}}

\begin{flushleft}
UDC 530.12; 531.51%
\end{flushleft}

\begin{center}
\textbf{COLLECTIVE EXCITATIONS OF QUANTUM UNIVERSE AND THE PROBLEM
OF DARK MATTER AND ENERGY\footnote{A talk given at the International
Conference ``Modern Problems of Theoretical Physics'' Dedicated to
the 90th anniversary of A.S. Davydov (December 9-15, 2002, Kiev, Ukraine)}}\\[0.5cm]
\textsf{V.E. Kuzmichev, V.V. Kuzmichev}\\[0.5cm]
\textit{Bogolyubov Institute for Theoretical Physics, \\ Nat.
Acad. of Sci. of Ukraine}\\
\textit{14b, Metrolohichna Str., 03143 Kiev, Ukraine}\\
\end{center}

\begin{abstract}
In the present report we make an attempt to give an answer to the
question of modern cosmology about the nature of the dark matter
and energy in our universe. On the basis of our quantum
cosmological approach, proposed in 1997-2002, we show that there
can be two types of collective states in the universe. One of them
relates to a gravitational field, another is connected with a
matter (scalar) field which fills the universe on all stages of
its evolution. The collective excitations of the scalar field
above its true vacuum reveal themselves mainly in the form of dark
matter and energy. Under the action of the gravitational forces
they decay and produce the non-baryonic dark matter, optically
bright and dark baryons. We have calculated the corresponding
energy densities which prove to be in good agreement with the data
from the recent observations.
\end{abstract}

\begin{center}
\textbf{1. Introduction}
\end{center}

Modern cosmology poses the principle question about the nature of
the components of our universe. The recent astrophysical
observations provide the strong empirical evidence that the total
energy density of the universe equals to $\Omega_{0} = 1$ to
within 10\% (in units of critical density) \cite{1,2}. At the same
time the mean matter density consistent with big-bang
nucleosynthesis is estimated by the value $\Omega_{M} \approx 0.3$
\cite{2}-\cite{4}. It is assumed that the density $\Omega_{M}$ is
contributed by the optically bright baryons ($\Omega_{b} \approx
0.005$) and dark matter which consists of the dark baryons
($\Omega_{db} \approx 0.04$) and matter of uncertain origin and
composition (known as cold dark matter, $\Omega_{cdm} \approx
0.3$). The contribution from neutrinos is negligible,
$\Omega_{\nu} \lesssim 0.003$. The lack of density $\Omega_{X}
\sim 0.7$, where $\Omega_{X} = \Omega_{0} - \Omega_{M}$, is
ascribed to so-called dark energy \cite{3}-\cite{6}. Its nature is
unknown and expected properties are unusual. This dark energy is
unobservable (in no way could it be detected in galaxies),
spatially homogeneous and, as it is expected, it has large,
negative pressure. The last property should guarantee an agreement
with the present-day accelerated expansion of the universe
observed in measurements of type Ia supernovae \cite{7,8}.

In this report we make an attempt to explain the matter structure
of our universe on the basis of the conjuncture that the
collective excitations of previously unknown type exist in it. At
the heart of this approach we put the quantum cosmology formulated
in Refs. \cite{9}-\cite{13}. It is shown that there are two types
of collective states in the quantum universe. One of them relates
to a gravitational field, another is connected with a matter
(scalar) field which fills the universe on all stages of its
evolution. The quantization of the fields which describe the
geometrical and matter properties of the universe is made. We
demonstrate that in the early epoch the non-zero energy of the
vacuum in the form of the primordial scalar field is a source of
the transitions of the quantum universe from one state to another.
During these transitions the number of the quanta of the
collective excitations of the gravitational field increases and
this growth manifests itself as an expansion of the universe. The
collective excitations of the scalar field above its true vacuum
reveal themselves mainly in the form of dark (nonluminous) matter
and energy. Under the action of the gravitational forces they
decay and produce the non-baryonic dark matter, optically bright
and dark baryons and leptons. For the state of the universe with
large number of the matter field quanta the total energy density,
density of (both optically bright and dark) baryons, and density
of non-baryonic dark matter are calculated. The theoretical values
prove to be in good agreement with the data from the recent
observations in our universe.

\textbf{On the system of units.} In this report we use
dimensionless variables. The modified Planck values are chosen as
units of length $l = \sqrt{2 G /(3 \pi)} = 0.74 \times 10^{-33}$
cm, mass/energy $m_{p} = \sqrt{3 \pi /(2 G)} = 2.65 \times
10^{19}$ GeV, and time $t_{p} = \sqrt{2 G /(3 \pi)} = 2.48 \times
10^{-44}$ sec, where $\hbar = c = 1$. The unit of density is
$\rho_{p} = 3 /(8 \pi G l^{2}) = \frac{9}{16}\,\rho_{pl}$, where
$\rho_{pl}$ is the standard Planck density. The scalar field
$\phi$ is measured in units $\phi_{p} = \sqrt{3 /(8 \pi G)} =
l \sqrt{\rho_{p}}$, while the potential $V(\phi)$ in $\rho_{p}$.\\

\newpage
\begin{center}
\textbf{2. Quantum equation of motion}
\end{center}

Let us consider a homogeneous, isotropic and closed universe
filled with primordial matter in the form of the uniform scalar
field $\phi$. The potential $V(\phi)$ of this field will be
regarded as a positive definite function of $\phi$. The quantum
analog of such universe is described by the equation \cite{9}
-\cite{13}
\begin{equation}\label{1}
   2\,i\, \partial _{T} \Psi = \left[ \partial _{a}^{2} -
  \frac{2}{a^{2}}\,\partial _{\phi }^{2} - U \right] \Psi \,,
\end{equation}
where its wavefunction $\Psi (a, \phi, T)$ depends on the scale
factor $a$ ($0 \leq a < \infty$), the scalar field $\phi$ ($-
\infty < \phi < \infty$), and the time variable $T$ ($- \infty < T
< \infty$). The last is related to the synchronous proper time $t$
by the differential equation $dt = a\,d T$ \cite{9,10}. The
functional
\begin{eqnarray}\label{2}
    U = a^{2} - a^{4}\,V(\phi )
\end{eqnarray}
plays the role of an effective potential of the interaction
between the gravitational and matter fields. It should be
emphasized that the equation (\ref{1}) is not an ordinary
Schr\"{o}dinger equation with a potential (\ref{2}). It is a
constraint equation for the wavefunction (see, e.g.,
\cite{Dirac,16}). The famous Wheeler-DeWitt equation for the
minisuperspace model (the wavefunction depends on the finite
number of variables) is the special case ($\partial_{T} \Psi = 0$)
of the more general equation (\ref{1}). The discussion of the
possible solutions of this equation, their interpretation, and the
corresponding references can be found in the Reviews \cite{17,18}.
In the region $a^{2} V
> 1$ the solutions of the Wheeler-DeWitt equation  and the
solutions of (\ref{1}) as well can be associated with the
exponential expansion of the universe \cite{9,13,22}.

A general solution of (\ref{1}) can be represented by a
superposition
\begin{equation}\label{3}
  \Psi (a, \phi , T) = \int_{- \infty}^{\infty }\!dE\,\mbox{e}^{\frac{i}{2} E T}\,
               C(E)\, \psi _{E}(a, \phi )
\end{equation}
of the states $\psi _{E}$ which satisfy a stationary equation
\begin{eqnarray}\label{4}
 \left( -\,\partial _{a}^{2} + \frac{2}{a^{2}}\,\partial _{\phi }^{2} +
             U - E  \right)  \psi _{E} = 0,
\end{eqnarray}
where $E$ is the eigenvalue. It has a physical dimension of
action, while here it is written in units $\hbar/2$. The function
$C(E)$ characterizes the distribution in $E$ of the states of the
universe at the instant $T = 0$.

Let us compare (\ref{1}) and (\ref{4}) with the corresponding
Schr\"{o}dinger equations. It is well known that when a
Hamiltonian does not depend explicitly on the time variable the
quantum system is described by the states with the time-dependence
in the form $\exp (-i \mathcal{E} t)$. The time variable $t$ and
the classical energy $\mathcal{E}$ can be related with each other
through the formal transformation $\mathcal{E} \rightarrow i\,
\partial _{t}$ \cite{14,15}. In the case of quantum cosmology
there is an analogous formal transformation between the classical
parameter $E$, which enters into the energy-momentum tensor of
radiation
\begin{eqnarray}\label{5}
  \tilde T^{0}_{0} & = & \frac{E}{a^{4}},\
 \tilde T^{1}_{1} = \tilde T^{2}_{2} = \tilde T^{3}_{3} =
 -\,\frac{E}{3\, a^{4}},\nonumber \\
 \tilde T^{\mu }_{\nu } & = & 0  \ \mbox{for} \ \mu \neq \nu,
\end{eqnarray}
and the time variable $T$, $E \rightarrow -\, 2\, i\, \partial
_{T}$ (minus originates from the specific character of the
gravitational problem, while the additional factor $2$ results
from the choice of units). This analogy demonstrates that the
role, played in ordinary quantum mechanics by the time variable
$t$, is assumed by the variable $T$. Thus in the framework of the
minisuperspace model the equation (\ref{1}) solves the problem of
correct definition of the time coordinate in quantum cosmology.
The role of time in quantum gravity and the difficulties with the
introduction of time variable suitable for the description of the
dynamics in quantum cosmology are elucidated in the Reviews
\cite{16,17}.

The effective potential $U$ has the form of barrier of the finite
width and height in the variable $a$. Therefore it is convenient
to represent the solution of (\ref{4}) in the form
\begin{equation}\label{6}
  \psi _{E} (a, \phi) = \int_{- \infty}^{\infty }\!d\epsilon\,
  \varphi _{\epsilon}(a,  \phi )\, f_{\epsilon} (\phi ; E),
\end{equation}
where the function $\varphi_{\epsilon}$ satisfies the equation
\begin{eqnarray}\label{7}
 \left( -\,\partial _{a}^{2} + U - \epsilon \right)  \varphi _{\epsilon} =
 0.
\end{eqnarray}
The eigenvalue $\epsilon$ and eigenfunction $\varphi_{\epsilon}$
depend parametrically on $\phi$. The functions
$\varphi_{\epsilon}$ describe the states of the continuous
spectrum. Using the explicit form of the solution of (\ref{7}) in
the region $a^{2} V \gg 1$ \cite{9,13}, it is easy to show that
the functions $\varphi_{\epsilon}$ can be normalized by the
condition
\begin{eqnarray}\label{8}
    \int_{0}^{\infty}\!da\ \varphi _{\epsilon}^{*} (a, \phi)\,
    \varphi _{\epsilon '} (a, \phi) = \delta (\epsilon - \epsilon
    ').
\end{eqnarray}
Then the function $f_{\epsilon}$ will satisfy the equation
\begin{eqnarray}\label{10}
    \partial _{\phi}^{2} f_{\epsilon} + \int_{-\infty}^{\infty}\!d\epsilon '
    K_{\epsilon \epsilon '}\, f_{\epsilon '} = 0,
\end{eqnarray}
where the kernel
\begin{eqnarray}\label{11}
    K_{\epsilon \epsilon '}(\phi; E) = \int_{0}^{\infty}\!da\ \varphi _{\epsilon}^{*}\,
    \partial_{\phi}^{2} \varphi_{\epsilon '}
\end{eqnarray}
\begin{eqnarray*}
     +  2\,\int_{0}^{\infty}\!da\
    \varphi _{\epsilon}^{*}\, \partial_{\phi} \varphi_{\epsilon '} \partial_{\phi}
     + \frac{1}{2}\,(\epsilon ' - E)\, \int_{0}^{\infty}\!da\ \varphi _{\epsilon}^{*}\,
     a^{2} \varphi_{\epsilon '}.\nonumber
\end{eqnarray*}

This equation can be reduced to a solvable form if one defines
concretely the problem. In order to find the function $\psi_{E}$,
the equation (\ref{4}) must be supplemented with the boundary
condition. The solutions which describe the quasistationary states
(with $\epsilon = \epsilon_{n} +i \,\Gamma_{n}$, where
$\epsilon_{n}$ and $\Gamma_{n}$ are the positions and widths of
the levels with the numbers $n = 0, 1, 2 \ldots$) are the most
interesting. The waves $\varphi_{\epsilon}$ outgoing toward
greater values of $a$ correspond to such states \cite{9,10,13}.

Just as in classical cosmology which uses a model of the slow-roll
scalar field \cite{18,19}, in quantum cosmology based on the
equation (\ref{1}) it makes sense to consider a scalar field
$\phi$ which slowly evolves into its true vacuum state,
$V(\phi_{vac}) = 0$, from some initial state $\phi =
\phi_{start}$, where $V(\phi_{start}) \sim \rho_{pl}$.

For the slow-roll potential $V$, when $\left|d \ln V / d \phi
\right|^{2} \ll 1$, the function $\varphi _{\epsilon }$ describes
the universe in the adiabatic approximation. In this case one can
separate the slow motion (with respect to the variable $\phi$)
from the rapid one (with respect to the variable $a$) in the
universe. This means that the universe expands (or contracts) more
rapidly than the state of matter has time to change.

The calculations demonstrate \cite{10,11,13} that the first ($n =
0$) quasistationary state appears when $V$ decreases to $V_{in} =
0.08 = 0.045 \rho_{pl}$. It has the parameters: $\epsilon_{0} =
2.62 = 1.31 \hbar, \ \Gamma_{0} = 0.31 = 1.25 \times 10^{43}\,
\mbox{s}^{-1}$. The lifetime of the universe in this state, $\tau
= 0.8 \times 10^{-43}$ s, is three times greater than the Planck
time $t_{p}$. The instant of the origination of the first
quasistationary state can be taken as a reference point of time
against the scale of $T$.

A further decrease of $V$ leads to an increase of the number of
quasistationary states of the universe. The levels which have
emerged earlier shift towards the oscillator values
$\epsilon_{n}^{0} = 4 n + 3$ which they would have in the limit $V
\rightarrow 0$ (see below), while their widths
\begin{equation}\label{12}
    \Gamma_{n} \approx \exp \left\{- 2 \!\int_{a_{1}}^{a_{2}} da \sqrt{U -
    \epsilon_{n}}\right\} \quad \mbox{at} \quad \Gamma_{n}\ll \epsilon_{n}
\end{equation}
decrease exponentially. Here $a_{1} < a_{2}$ are the turning
points specified by the condition $U = \epsilon_{n}$. As a result
the lifetime of the universe in one of such states is many orders
greater than the Planck time and at $V \sim 10^{-122}$ it reaches
the values $\tau \sim 10^{61} \sim 10^{17}$ s comparable with the
age of our universe.\\

\begin{center}
\textbf{3. Collective states of the gravitational field}
\end{center}

Within the lifetime the state of the universe can be considered
with a high accuracy as a stationary state which takes the place
of a quasistationary one when its width becomes zero \cite{20}.
Taking into account that such states emerge at $V \ll 1$, while in
the prebarrier region ($a < a_{1}$) $a^{2} V < 1$ always, the
solutions of (\ref{7}) can be written in the form of expansion in
powers of $V$ on the interval $\Delta \phi = |\phi_{vac} -
\phi_{start}|$. We have
\begin{eqnarray*}
\varphi_{n} = \left.|n\right\rangle
  - \frac{V}{4}\left[\frac{1}{8}\sqrt{N(N-1)(N-2)(N-3)}
\left.|n-2\right\rangle \right.\nonumber\\
  + \sqrt{N(N-1)}\left(N- \frac{1}{2}\right)\left.|n-1\right\rangle
  \nonumber \\
- \sqrt{(N+1)(N+2)}\left(N+\frac{3}{2}\right)\left.|n+1\right\rangle \nonumber \\
 - \left.
\frac{1}{8}\sqrt{(N+1)(N+2)(N+3)(N+4)}\left.|n+2\right\rangle\right]\nonumber
\end{eqnarray*}
\begin{eqnarray}\label{13}
-\ O(V^{2})
\end{eqnarray}
for the wavefunction and
\begin{eqnarray}\label{14}
\epsilon_{n}  = \epsilon_{n}^{0} - \frac{3}{4}\,V\left[2N(N+1) +
1\right] - O(V^{2})
\end{eqnarray}
for the position of the level, where $N = 2n + 1$. Here $|n
\rangle$ is the eigenfunction, $\epsilon_{n}^{0} = 2 N + 1$ is the
eigenvalue of the equation for an isotropic oscillator with zero
orbital angular momentum
\begin{equation}\label{15}
    \left( - \partial_{a}^{2} + a^{2} - \epsilon_{n}^{0}\right)\left.|n\right\rangle =
    0.
\end{equation}

The wavefunction $\langle a|n \rangle$ describes the geometrical
properties of the universe as a whole. Since the gravitational
field is considered as a variation of space-time metric \cite{21},
then this wavefunction also characterizes the quantum properties
of the gravitational field which is considered as an aggregate of
the elementary excitations of a quantum oscillator (\ref{15}). It
should be emphasized that in this approach the role of the dynamic
variables is taken by the variables $(a, \phi)$ of the
minisuperspace and after the quantization such elementary
excitations cannot be identified with gravitons. Introducing the
operators
\begin{equation}\label{16}
A^{\dag}= \frac{1}{\sqrt{2}}\,(a - \partial_{a}), \quad
 A = \frac{1}{\sqrt{2}}\,(a + \partial_{a}),
\end{equation}
the state $|n \rangle$ can be represented in the form
\begin{eqnarray}\label{17}
 |n \rangle  =
 \frac{1}{\sqrt{N!}}\,(A^{\dag})^{N}\,|vac \rangle, \quad
  A\,|vac \rangle  = 0, \nonumber\\ |vac \rangle =
 \left(\frac{4}{\pi}\right)^{1/4}\exp \left\{-
 \frac{a^{2}}{2}\right\}.
\end{eqnarray}
Since $A^{\dag}$ and $A$ satisfy the ordinary canonical
commutation relations, $[A, A^{\dag}] = 1,\ [A,A] = [A^{\dag},
A^{\dag}] = 0$, then one can interpret them as the operators for
the creation and annihilation of quanta of the collective
excitations of the gravitational field (in the sense indicated
above). We shall call it g-quantum. The integer $N$ gives the
number of g-quanta in the n-th state of the gravitational field.

For clearness we rewrite (\ref{15}) in the physical units using
for this purpose the Hamiltonian from \cite{9},
\begin{eqnarray}\label{18}
\left( - \frac{\hbar^{2}}{2 m_{p}}\,\partial_{R}^{2} +
\frac{k_{p}}{2}\, R^{2}\right) \psi_{n}(R) = E_{n}\psi_{n}(R),
\end{eqnarray}
where $R = l a$ is a ``radius'' of the curved universe, while
$k_{p}= (m_{p}\,c^{2})^{3}\, (\hbar\,c)^{-2}$ can be called a
``stiffness coefficient of space (or gravitational field)''. Its
numerical value is $k_{p} = 4.79 \times 10^{85} \,
\mbox{GeV}/\mbox{cm}^{2} = 0.76 \times 10^{83} \,
\mbox{g}/\mbox{s}^{2}$.

According to (\ref{18}) the states (\ref{17}) can be interpreted
as those which emerge as a result of motion of some imaginary
particle with the mass $m_{p}$ and zero orbital angular momentum
in the field with the potential energy $U(R) = k_{p} R^{2} /2$.
This motion causes the equidistant spectrum of energy $E_{n}=
\hbar\,\omega_{p}\left(N + \frac{1}{2}\right) =
2\,\hbar\,\omega_{p}\left(n + \frac{3}{4}\right)$, where
$\hbar\,\omega_{p} = m_{p}\,c^{2}$ is the energy of g-quantum,
$\omega_{p} = t_{p}^{-1}$ is the oscillation frequency.

In the quantum cosmological system of the most general type which
is a superposition of the waves $\varphi_{\epsilon}$ incident upon
the barrier and scattered by the barrier \cite{9,10,13}, it occurs
the formation of the fundamentally new state as a result of a slow
decreasing of the potential $V$ (vacuum energy density
\cite{17,18}). When the potential $V$ reaches the value $V_{in}$
the wave $\varphi_{\epsilon}$ with $\epsilon \approx \epsilon_{0}$
penetrates into the prebarrier region \cite{13} and it results in
the transition of the cosmological system to a quasistationary
state. In this state the system is characterized by the
expectation values of the scale factor $\langle \varphi_{0} | a |
\varphi_{0} \rangle = 1.25 = 0.93 \times 10^{-33} \, \mbox{cm}$
and total energy density $\rho_{n=0} = 1.16 = 0.65 \rho_{pl} =
3.35 \times 10^{93} \, \mbox{g}/\mbox{cm}^{3}$ \cite{10,13}. These
parameters are of Planck scale. Such a new formation has a
lifetime which exceeds the Planck time and one can consider it as
a quantum cosmological system with well-defined physical
properties. We call it the quantum universe. Such universe can
evolve by means of a change of its quantum state. In every quantum
state it can be characterized by the energy density, observed
dimensions, lifetime (age), proper dimensions of non-homogeneities
of the matter density, amplitude of fluctuation of radiation
temperature, power spectrum of density perturbations, angular
structure of the radiation anisotropies, deceleration parameter,
total entropy, and others \cite{13}.

The interaction of the gravitational field with a vacuum which has
a non-zero energy density ($V(\phi) \neq 0$) in the range $\Delta
\phi$ results in the fact that the wavefunction (\ref{13}) is a
superposition of the states $|n \rangle, \ |n \pm 1\rangle$ and
$|n \pm 2 \rangle$. From (\ref{13}) and (\ref{14}) it follows that
the quantum universe can be characterized by a quantum number $n$.

The physical state $|n \rangle$ (\ref{17}) is chosen so that $|0
\rangle$ describes an initial state of the gravitational field.
From the point of view of the occupation number representation
there is only one g-quantum in such state. When the gravitational
field transits from the state $|n \rangle$ into the neighbouring
one $|n + 1 \rangle$ two g-quanta are created, while the
corresponding energy increases by $2\, \hbar\, \omega_{p}$. In the
inverse transition two g-quanta are absorbed and the energy
decreases. The transitions with the creation/absorbation of one
g-quantum are forbidden. The vacuum state $|vac \rangle$ in
(\ref{17}) describes the universe without any g-quantum.

The g-quanta are bosons. Since the probability of the creation of
a boson per unit time grows with the increase in number of bosons
in a given state \cite{14}, then an analogous effect must reveal
itself in the quantum universe during the creation of g-quanta. As
$\langle a \rangle \sim \sqrt{(N + 1)/2}$, where the brackets
denote the averaging over the states $\varphi_{n}$, then the
growth in number $N$ of g-quanta (or number of levels $n$) means
the increase in expectation value of the scale factor. In other
words the expansion of the universe reflects the fact of the
creation of g-quanta (under the transition to a higher level). The
increase in probability of the creation of these quanta per unit
time leads to the accelerated expansion of the universe. This
phenomenon becomes appreciable only when $n$ reaches the large
values. The observations of type Ia supernovae provide the
evidence that today our universe is expanding with acceleration
\cite{4,7,8}. The qualitative explanation of this phenomenon
mentioned above is confirmed by the concrete quantitative
calculations of the deceleration parameter \cite{13}. On this
point the theory is in good agreement with the measurements of
type Ia supernovae.

When the potential $V$ decreases to the value $V \ll 0.1$ the
number of available states of the universe increases up to $n \gg
1$. By the moment when the scalar field will roll in the location
where $V(\phi_{vac}) = 0$ the universe can be found in the state
with $n \gg 1$. This can occur because the emergence of new
quantum levels and the (exponential) decrease in width of old ones
result in the appearance of competition between the tunneling
through the barrier $U$ and allowed transitions between the
states, $n \rightarrow n, \ n \pm 1, \ n \pm 2$. A comparison
between these processes demonstrates \cite{9,10,13} that the
transition $n \rightarrow n + 1$ with creation of g-quanta is more
probable than any other allowed transitions or decay. The vacuum
energy  of the early universe originally stored by the field
$\phi$ with a potential $V(\phi_{start})$ is a source
of transitions.\\

\begin{center}
\textbf{4. Collective states of the matter field}
\end{center}

According to accepted model the scalar field $\phi$ descends to
the state with zero energy density, $V(\phi_{vac}) = 0$. Then the
field $\phi$ begins to oscillate with a small amplitude about
equilibrium vacuum value $\phi_{vac}$. We shall describe these new
states of the quantum universe assuming that by the moment when
the field $\phi$ reaches the value $\phi_{vac}$ the universe
transits to the state with $n \gg 1$. This means that the
``radius'' of the universe has become $\langle a \rangle > 10\,
l$. The wavefunction of the universe in the state with a given $n$
takes the following form up to the terms $\sim O(V^{2})$
\begin{equation}\label{19}
    \psi_{n} (a, \phi) = \langle a | n \rangle \,f_{n}(\phi; E)
    \quad \mbox{at} \quad n \gg 1,
\end{equation}
where the function $f_{n}$ satisfies the equation
\begin{equation}\label{20}
    \left[\partial_{x}^{2} + z - V(x) \right] f_{n} = 0.
\end{equation}
Here $x = \sqrt{m/2}\,(2N)^{3/4}\,(\phi - \phi_{vac})$
characterizes the deviation of the field $\phi$, $z =
(\sqrt{2N}/m)\,\left(1 - E/(2N)\right),$ $V(x) =
(2N)^{3/2}\,V(\phi)/m$, and $m$ is a dimensionless parameter which
it is convenient to choose as $m^{2} = \left[\partial_{\phi}^{2}
V(\phi)\right]_{\phi_{vac}}$. The formulae (\ref{19}) and
(\ref{20}) follow from (\ref{6}), (\ref{10}), and (\ref{13}) if
one takes into account resonance behaviour of the wave
$\varphi_{\epsilon}$ at $\epsilon = \epsilon_{n}$ \cite{11,13}.

Since $\langle a \rangle = \sqrt{N/2}$, where averaging was
performed over the states (\ref{19}), then $V(x)$ is a potential
energy of the scalar field contained in the universe with the
volume $\sim \langle a \rangle^{3}$. Expanding $V(x)$ in powers of
$x$ we obtain
\begin{equation}\label{21}
    V(x) = x^{2} + \alpha\,x^{3} + \beta\,x^{4} + \ldots,
\end{equation}
where the parameters
\begin{equation}\label{22}
    \alpha =
    \frac{\sqrt{2}}{3}\,\frac{\lambda}{m^{5/2}}\,\frac{1}{(2N)^{3/4}}\,,
    \
    \beta = \frac{1}{6}\,\frac{\nu}{m^{3}}\,\frac{1}{(2N)^{3/2}}\,,
\end{equation}
and $\lambda = \left[\partial_{\phi}^{3} V(\phi)
\right]_{\phi_{vac}}, \ \nu = \left[\partial_{\phi}^{4} V(\phi)
\right]_{\phi_{vac}}$. Since $N \gg 1$, then the coefficients
$|\alpha| \ll 1$ and $|\beta| \ll 1$ even at $m^{2} \sim \lambda
\sim \nu$. Therefore the equation (\ref{20}) can be solved using
the perturbation theory for stationary problems with a discrete
spectrum. We take for the state of the unperturbed problem the
state of the harmonic oscillator with the equation of motion
\begin{equation}\label{23}
    \left[\partial_{x}^{2} + z^{0} - x^{2} \right] f_{n}^{0} = 0.
\end{equation}
In the occupation number representation one can write
\begin{eqnarray}\label{24}
 f_{ns}^{0} = \frac{1}{\sqrt{s!}}\,(B_{n}^{\dag})^{s} f_{n0}^{0},
 \quad B_{n}\,f_{n0}^{0} = 0, \nonumber \\ f_{n0}^{0} (x) =
 \left(\frac{1}{\pi}\right)^{1/4}\exp \left\{-
 \frac{x^{2}}{2}\right\}
\end{eqnarray}
with $z^{0} = 2s + 1$, where $s = 0, 1, 2 \ldots$, and
\begin{equation}\label{25}
B_{n}^{\dag}= \frac{1}{\sqrt{2}}\,(x - \partial_{x}), \qquad
 B_{n} = \frac{1}{\sqrt{2}}\,(x + \partial_{x}).
\end{equation}
Here $B_{n}^{\dag}$ ($B_{n}$) can be interpreted as the creation
(annihilation) operator which increases (decreases) the number of
quanta of the collective excitations of the scalar field in the
universe in the n-th state. We shall call them the matter quanta.
The variable $s$ is the number of matter quanta in the state $n
\gg 1$. It can be considered as an additional quantum number.

Using (\ref{20}), (\ref{21}) and (\ref{24}) we obtain
\begin{equation}\label{26}
    z = 2 s + 1 + \Delta z,
\end{equation}
where
\begin{eqnarray}\label{27}
    \Delta z & = & \frac{3}{2}\,\beta \left(s^{2} + s +
    \frac{1}{2}\right) - \frac{15}{8}\,\alpha^{2} \left(s^{2} + s +
    \frac{11}{30}\right) \nonumber\\
    & - & \frac{\beta^{2}}{16} \left(34 s^{3} + 51 s^{2} + 59 s +
    21\right)
\end{eqnarray}
takes into account a self-action of matter quanta. In order to
determine the physical meaning of the quantities which enter the
equation (\ref{20}) with the potential (\ref{21}) we rewrite it in
the ordinary physical units,
\begin{eqnarray}\label{28}
\left( - \frac{\hbar^{2}}{2 \mu}\,\partial_{r}^{2} +
\frac{1}{2}\,\mu\, \omega^{2} r^{2} + {\cal E}_{1}\,\left(
\frac{r}{l}\right)^{3} + {\cal E}_{2}\,\left(
\frac{r}{l}\right)^{4} \right. \nonumber\\
\left. + \ldots -
{\cal E} \right) f_{n}(r) = 0,
\end{eqnarray}
where we denote $\mu = m_{p}\,m^{-1},\ r = l\,x,\ \omega =
m\,t_{p}^{-1},\ {\cal E}_{1} = m \, m_{p}\, c^{2}\, \alpha / 2, \
{\cal E}_{2} = m \, m_{p}\, c^{2}\, \beta / 2,$ ${\cal E} = m \,
m_{p}\, c^{2}\, z / 2$. Here $l = \sqrt{\hbar/(\mu \omega)}$ is
the Planck length. According to (\ref{28}) an imaginary particle
with a mass $\mu$ performs the anharmonic oscillations and
generates the energy spectrum
\begin{equation}\label{29}
    {\cal E} = \hbar\,\omega \left(s + \frac{1}{2} + \frac{\Delta
    z}{2}\right),
\end{equation}
where $\hbar\, \omega = m\,m_{p}\,c^{2}$ is the energy of matter
quantum, and $m$ can be interpreted as its mass (in units
$m_{p}$). The quantity
\begin{equation}\label{29a}
    M = m \,\left(s + \frac{1}{2}\right) + \Delta M,
\end{equation}
where $\Delta M = m\,\Delta z/2$, is a mass of the universe with
$s$ matter quanta. Since we are interested in large values of $s$,
then for the estimation of $\Delta M$ let us assign the values $s
\sim 10^{80}$ and $n \sim 10^{122}$. These parameters describe our
universe, where $s$ is equal to the equivalent number of baryons
in it and $\langle a \rangle \sim 10^{28}$ cm is a size of its
observed part \cite{11,12}. In this case $\Delta M \sim
O\left((\nu/m^{2}) 10^{-24},\ (\lambda/m^{2})^{2}
10^{-24}\right)$. Hence when the number of the matter quanta
becomes very large their self-action can be a fortiori neglected.\\

\begin{center}
\textbf{5. The matter structure of the quantum universe}
\end{center}

In the early epoch (on the interval $\Delta \phi$) there are no
matter in the ordinary sense in the universe, since the field
$\phi$ is only a form of existence of the vacuum, while the vacuum
energy density is decreasing with time. In spite of the fact that
$V(\phi_{vac}) = 0$, the average value $\langle V \rangle \neq 0$.
It contributes to the vacuum energy density in the epoch when the
matter quanta are created and determines the cosmological constant
$\Lambda = 3 \, \langle V \rangle$. The universe is filling with
matter in the form of aggregate of quanta of the collective
excitations of the primordial scalar field.

Let us consider the possible matter structure of the universe from
the point of view of quantum cosmology. For this purpose we change
from the quantum equation (\ref{4}) to the corresponding Einstein
equation for homogeneous, isotropic, and closed universe filled
with the uniform matter and radiation. Averaging over the states
$\psi_{E}$ we obtain the Einstein-Friedmann equation in terms of
the average values
\begin{equation}\label{30}
   \left(\frac{\partial_{t} \langle a \rangle}{\langle a \rangle}
    \right)^{2} = \langle \rho \rangle - \frac{1}{\langle a
    \rangle^{2}}\,,
\end{equation}
where we have neglected the dispersion, $\langle a^{2} \rangle
\sim \langle a \rangle ^{2}$, and $\langle a^{6} \rangle \sim
\langle a \rangle ^{6}$. The average value
\begin{equation}\label{31}
    \langle \rho \rangle = \langle V \rangle + \frac{2}{\langle a
    \rangle^{6}}\,\langle - \partial_{\phi}^{2} \rangle + \frac{E}{\langle a \rangle^{4}}
\end{equation}
is a total energy density in some fixed instant of time with the
Hubble constant $H_{0} = \partial_{t} \langle a \rangle / \langle
a \rangle$. The first term is the energy density of the vacuum
with the equation of state $p_{v} = - \rho_{v} = - \langle V
\rangle$, the second term is the matter energy density, while the
last describes the contribution of the radiation.

Using the wavefunction (\ref{19}) and passing in (\ref{17}) and
(\ref{24}) to the limit of large quantum numbers, for the energy
densities of the vacuum $\Omega_{v}$ and matter $\Omega_{qm}$ we
obtain (in units of critical density $\rho_{c} = H_{0}^{2}$)
\begin{eqnarray}\label{32}
 \Omega_{v} =  \frac{M}{12\langle a
 \rangle^{3}\,H_{0}^{2}}\,,\quad \Omega_{qm}= \frac{16\, M}{\langle a
 \rangle^{3}\,H_{0}^{2}}\,,
\end{eqnarray}
where $M = m \left(s + \frac{1}{2}\right)$ is the mass of the
universe with ``radius'' $\langle a \rangle = \sqrt{N/2}$. From
the definition of $z$ in (\ref{20}) it follows that $\langle a
\rangle = M + E/(4 \langle a \rangle)$. For the matter-dominant
era, $M \gg E/(4 \langle a \rangle)$, from (\ref{30}) and
(\ref{32}) we find
\begin{equation}\label{33}
    \Omega = 1.066, \quad \Omega_{v} = 0.006, \quad \Omega_{qm} =
    1.060,
\end{equation}
where $\Omega = \langle \rho \rangle H_{0}^{-2}$. From (\ref{33})
it follows that the quantum universe in all states with $n \gg 1$
and $s \gg 1$ looks like spatially flat. A main contribution to
the energy density of the universe is made by the collective
excitations of the scalar field above its true vacuum. The total
density $\Omega$ can be compared with the present-day density of
our universe, $\Omega_{0} = 1 \pm 0.12$ \cite{1} or $\Omega_{0} =
1.025 \pm 0.075$ \cite{2}.

The matter quanta are bosons. They have non-zero mass/energy,
while their state according to (\ref{24}) depends on the state of
the gravitational field. This means that the matter quanta are
subject to the action of gravity. Due to this fact they can decay
into the real particles (for example, baryons and leptons) that
have to be present in the universe in small amount (because of the
weakness of the gravitational interaction). The main contribution
to the energy density will still be made by the matter quanta.

In order to make a numerical estimate we consider the matter
quanta ($\phi$) decay scheme
\begin{equation}\label{34}
    \phi \rightarrow b + n \rightarrow \phi' + \nu + p + e^{-} +
    \bar{\nu}.
\end{equation}
The daughter quantum $b$ decays into the quantum $\phi'$ of the
residual excitation, which reveals itself in the universe in the
form of the non-baryonic dark matter, and neutrino $\nu$, which
takes away the spin and main part of the energy of the kinetic
motions of the particles after the decay of the quantum $\phi$.
The density of (optically bright and dark) baryons equals to
\begin{eqnarray}\label{35}
 \Omega_{B}= \frac{16\, \bar{s}\, m_{proton}}{\langle a
 \rangle^{3}\,H_{0}^{2}}\,,
\end{eqnarray}
where $m_{proton} = 0.938$ GeV is a proton mass and $\bar{s}$ is
an average number of the quanta $\phi$ which decay during the time
interval $\Delta t$. Taking into account (\ref{32}) we find
\begin{equation}\label{36}
    \Omega_{B} = \Omega_{qm}\,{\sqrt{\frac{3 \pi}{2 g}}}
    \ \frac{m_{proton}}{m_{p}}
    \left(1 - \exp \left\{-\Gamma_{tot}\, \Delta t\right\} \right),
\end{equation}
where $\Gamma_{tot}$ is the mean probability of the decay on the
time interval $\Delta t$, $g = G\,m^{2}$ is the gravitational
coupling constant for the quantum $\phi$ with mass $m$.

It is obvious that calculating $\Gamma_{tot}$ one must take into
account both vertices of the decay (\ref{34}) with the same
accuracy. On the order of magnitude, the correct result can be
obtained in a first approximation which parametrizes the vertices
by the corresponding coupling constants. Then assuming that the
decay $\phi \rightarrow \phi'\,\nu\,n$ is caused by the action of
the gravitational forces, for the mean probability of the decay we
find a simple expression
\begin{equation}\label{37}
    \Gamma_{tot} = \alpha\,g\,\Delta m\,,
\end{equation}
where $\alpha$ is a fine-structure constant, $\Delta m = 1.293$
MeV is a difference in masses of neutron and proton. Since we are
interested in matter density in the universe today, then we choose
$\Delta t$ equal to the age of the universe, $\Delta t = 14$ Gyr
\cite{3}. Then substituting (\ref{37}) in (\ref{36}) and using the
numerical values of the parameters we find the dependence of the
density $\Omega_{B}$ on the coupling constant $g$. The function
$\Omega_{B}(g)$ obtained in such approach vanishes at $g = 0$ and
tends to zero as $g^{-1/2}$ at $g \rightarrow \infty$. It has one
maximum,
\begin{equation}\label{38}
    \Omega_{B} = 0.128 \quad \mbox{at} \quad g = 20 \times
    10^{-38}.
\end{equation}
The mass of matter quantum $\phi$ corresponding to such coupling
constant is equal to $m = 5.459$ GeV. The density of the
non-baryonic dark matter is
\begin{equation}\label{39}
    \Omega_{\bullet} = \Omega_{B} \, \frac{m_{\phi'}}{m_{proton}}\,,
\end{equation}
where, according to (\ref{34}), $m_{\phi'} = m - m_{n} -
\overline{m}_{\nu} - Q$ is a mass of the quantum $\phi'$,
$\overline{m}_{\nu} \equiv m_{\nu}^{2}/(2\,p_{\nu})$, $m_{\nu}$
and $p_{\nu}$ are the neutrino mass and momentum, $Q$ is the
energy of the relative motion of all particles. Since the
contribution from $\Omega_{\bullet}$ to the matter density
$\Omega_{M}$ of our universe is not at least smaller than
$\Omega_{B}$ \cite{2}-\cite{4}, then the mass $m_{\phi'}$ can
possess the values within the limits $0.938\,\mbox{GeV} \leq
m_{\phi'} \leq 4.519\,\mbox{GeV}$ depending on the value of $Q$.
Such particles are non-relativistic and non-baryonic dark matter
is classified as cold. Hence we find
\begin{equation}\label{40}
    0.128 \leq \Omega_{\bullet} \leq 0.617\,.
\end{equation}
Correspondingly the total matter density $\Omega_{M} = \Omega_{B}
+ \Omega_{\bullet}$ can possess any values within the limits
\begin{equation}\label{41}
    0.256 \leq \Omega_{M} \leq 0.745\,.
\end{equation}
A spread in theoretical values of the densities $\Omega_{\bullet}$
and $\Omega_{M}$ is caused by the fact that the energy $Q$ is an
undefined parameter of the theory. With the regard for this remark
one can compare (\ref{41}) with the matter density of our universe
\cite{3}: $0.2 \lesssim \Omega_{M}^{exp} \lesssim 0.4$, where a
spread in values is related with inaccuracy of measurements. We
accept as an illustration the value $\Omega_{M} \approx 0.3$ which
is in good agreement with theoretical calculation (\ref{41}). Then
we find $\Omega_{\bullet} \approx 0.172,\ m_{\phi'} \approx 1.260$
GeV and $Q \approx 3.259$ GeV. Such kinetic energy corresponds to
the temperature $T \approx 3.78 \times 10^{13}$ K. This
temperature can be in the primary plasma in the early universe at
$\Delta t \sim 10^{-7}$ s.

According to (\ref{33}) and (\ref{41}) the residual density
$\Omega_{X} = \Omega_{qm} - \Omega_{M}$ can possess the values
within the limits
\begin{equation}\label{42}
    0.315 \leq \Omega_{X} \leq 0.804\,.
\end{equation}
The observations in our universe give the restriction $0.6
\lesssim \Omega_{X}^{exp} \lesssim 0.8$ \cite{3}. These values
agree with (\ref{42}), where, however, a spread in values is
related with the uncertainty in $Q$ as in (\ref{41}).

In our approach according to (\ref{31}), (\ref{32}), (\ref{35}),
and (\ref{39}) the residual density $\Omega_{X}$ has only
dynamical nature and it can be attributed to the optically dark
(nonluminous) energy.

In conclusion we note that the present-day density of optically
bright and dark baryons in our universe is estimated as
$\Omega_{B}^{exp} = \Omega_{b} + \Omega_{db} \sim 0.045$
\cite{2}-\cite{4}. This value does not contradict with (\ref{38}),
since the latter determines the maximum possible baryon matter
density.

The experimental value of the cold dark matter density
$\Omega_{cdm} \sim 0.3$ falls within the limits of its theoretical
counterpart (\ref{40}). The density of the optically bright
baryons can be estimated as $\Omega_{*} =
\frac{1}{16}\,\Omega_{B}$. Then according to (\ref{38}) the
contribution from the optically bright baryons must not exceed
$\Omega_{*} \sim 0.008$. This value is in good agreement with the
evidence from the bright stars observations $\Omega_{*} \sim
0.005$ \cite{4,23}.

\end{document}